\documentclass[%
 reprint,
 amsmath,amssymb,
 aps,
]{revtex4-2}

\usepackage{graphicx}% Include figure files
\usepackage{bm}% bold math
\usepackage{dcolumn}% Align table columns on decimal point
\usepackage{float}
\usepackage{caption}
\usepackage{subfigure}
\usepackage{amsmath}
\usepackage{mathrsfs}
\usepackage{amssymb}
\usepackage{hyperref}% add hypertext capabilities
\usepackage{xcolor}
\begin{document}

\preprint{}

\title{Non-equilibrium ensemble theory for thermal transport in anharmonic crystals}% Force line breaks with \\
%\thanks{BB}%
\author{Li Wan}
%\altaffiliation[Also at ]{}%Lines break automatically or can be forced with \\
\email{lwan@wzu.edu.cn}
\affiliation{Department of Physics, Wenzhou University, Wenzhou 325035, P. R. China}%


\begin{abstract}
We propose an ensemble theory for the non-equilibrium statistics to study the thermal transport in anharmonic crystals. In the theory, lattice vibrations of the crystals are quantized by local Bosons(LBs), instead of Phonons as usually used for the thermal transport. LBs are driven by the temperature gradient and move from atom to atom in the crystals. Based on the LBs, anharmonic interactions between atoms in the crystals can be fully considered. To demonstrate our theory, we study the thermal transport in an atomic chain with a temperature drop applied on the two ends of the chain. We observe a Rabi-like oscillation in the transport of the LBs, from which we define the thermal current to get the thermal conductivity of the chain. Results show that the thermal conductivity is enhanced slowly with the increasing of the anharmonic interaction and decreases rapidly if the anharmonic interaction is increased further. In the present study, we only focus on the steady state, and the fluctuations of the thermal currents are not considered. 
\end{abstract}

\maketitle

%\tableofcontents

\section{introduction}
In non-metal crystals, the thermal transport is realized through the lattice vibrations~\cite{Born54}. To understand the thermal transport microscopically, the atomic interactions in the lattice vibrations need to be clarified in details~\cite{Mons13}. When the atomic interaction in one crystal is harmonic, the equations of motion for the lattice dynamics can be solved analytically. If weak anharmonicity is introduced in the atomic interaction, high-order terms in the potential expansion are considered as a small perturbation of the harmonic potential~\cite{Wall72,Ipato71,Sriva90}. However, when the anharmonicity dominates the atomic interaction, a general theory based on non-perturbation  of the full anharmonic potential is required for the thermal transport in the crystal. In this study, we propose such a general theory based on the non-equilibrium statistics.\\

The thermal transport can be investigated in the real space by using the molecular dynamics(MD) simulations~\cite{Frenk02}. The MD simulations are carried out for the crystals at the equilibrium state and can be used to calculate the correlations of the heat flux. The heat flux is defined as the time derivative of the product of the displacement and the energy of the atoms~\cite{McG06,Sche02,Hardy63,Volz99,Dong01, Turney09,English09}. The thermal conductivity is proportional to the time integral of the correlation function of the heat flux according to the Green-Kubo(GK) formalism~\cite{Green54,Kubo57}. The MD simulations are very flexible since the anharmonic interactions of the atoms in the crystals can be fully considered. The MD simulation combined with the large deviation theory is also a powerful tool to investigate the fluctuations of the thermal current in the real space~\cite{Ray19}.  However, the MD simulations need to record the trajectories of all the atoms in the real space for the GK formalism, and require intensive computational loads. In order to get the atomic interactions accurately, \textit{ab initio} MD simulations are applied, which normally can be carried out with very short simulation periods and very small simulation cells~\cite{Marco16,Carbogno17,Kang17,Kinaci12,Tse18}. Besides the MD simulations combined with the GK formalism at the equilibrium state, MD simulations can be carried out on the crystals directly with the temperature gradient applied~\cite{Turney09}. Similarly, the direct MD simulations still need intensive computational loads.  \\
 
The thermal transport can also be studied in the Phonon space. As is widely used, the collective vibrations of atoms in crystals are quantized by Phonons~\cite{Born54,Wall72,Ipato71,Sriva90}.  Phonons with momentum and energy are the quasi-particles for the thermal transport in the crystals. The relation between the frequency and the wave vector of the Phonons is the Phonon Dispersion Relation(PDR). In harmonic crystals where the interactions between atoms are harmonic, PDRs are lines. The derivative of the frequency with respect to the wave vector along the PDR lines defines the group velocity of Phonons. Based on the motion of Phonons, thermal conductivity calculated in harmonic crystals is infinite, which is wrong in reality. Thus, anharmonicity has to be introduced in the atomic potentials of the crystals to induce multiphonon scattering. In this way, the thermal conductivity obtained is finite. The multiphonon scattering includes the Normal(N) and the Umklapp(U) processes~\cite{Born54,Wall72,Ipato71,Sriva90}. After the introduction of the anharmonic interactions, the PDR lines are broadened and the group velocity defined by the derivatives along the PDR lines then is approximate.\\

In practice, Phonon Boltzmann Transport Equation(PBTE) has been used to calculate the thermal conductivity for the crystals~\cite{Pei29,Ziman60}. PBTE assumes that the anharmonic interaction between atoms is weak, so that the Phonon number of each mode follows the Bose-Einstein distribution at equilibrium. In the calculations by PBTE, Phonons move with the group velocities defined from the PDR for each mode. And the Phonon scatterings induced by the anharmonicity of atomic interactions, defects and boundaries are transformed to be the Phonon life time. The calculations of the Phonon life time can be done through two methods, the theory of lattice dynamics or the MD simulations~\cite{Mara62,Thomas10}.
As we have mentioned, the PBTE is valid only when the anharmonic interaction is weak. If the anharmonic interaction is strong, the Bose-Einstein distribution of Phonons will be violated and the PDR lines are broadened as well, even across each other~\cite{Simon19}. In this case, the Phonon is not well defined. Additionally, the anharmonic interaction induces multiphonon scatterings, which makes the computation of the Phonon life time very difficult. Thus, in anharmonic crystals, the accuracy of PBTE is in question and the definition of Phonon as the plane wave for the lattice vibrations may fail.\\

Open quantum theory(OQT) is a different way to study the thermal transport of the crystals~\cite{Gardiner00,Weimer21,Breuer16, Dhar08}. In the OQT, the crystals are connected to various thermal baths which are set at different temperatures. The temperature drops of the baths applied on the crystals drive the thermal energy to move in the crystals, which contributes to the thermal transport. The full Hamiltonian for the whole system contains not only the crystals but also the baths. However, it needs a large computational cost to solve the quantum equations with the full Hamiltonian. The OQT can be reduced to be stochastic dynamics, such as quantum langevin equation or Fokker-planck equation, if the Markovin approximation is applied~\cite{Wan17,Zeng19}. It is still very difficult to solve the equations of the stochastic dynamics if the anharmonicity is involved in the atomic interaction.\\

In this study, we quantize the atomic vibrations by local Bosons. Each atom in a crystal vibrating around its equilibrium position is considered as a quantum oscillator. Each quantum oscillator stimulates Bosons on its own local site. The Bosons are on site locally and different from Phonons which are for the collective lattice vibrations. The local Bosons can move from atom to atom in the lattice of the crystal if they are driven by the temperature gradient applied in the crystal. After introducing the local Bosons, we propose an ensemble theory for the local Bosons in the non-equilibrium state. Based on the ensemble theory, the thermal current is defined to study the thermal transport. In this theory, the anharmonic interactions between atoms can be fully considered. 

\section{theory}
To demonstrate our theory, we consider an atomic chain as an example for the crystal. The atomic chain has $N$ identical atoms indexed from $1$ to $N$. The atoms are arranged along the chain periodically. The lattice parameter of the chain is $l$, which should be determined by minimizing the potential energy between the atoms. The length of the chain is $L=l(N-1)$, and the mass of one atom is denoted by $M$. The theory of lattice dynamics tells us that the atoms vibrate around their own equilibrium positions. We denote the equilibrium position of the $j$-th atom by $R_j$ and the displacement of the $j$-th atom away from $R_j$ by $r_j$ for the vibration.\\

The Hamiltonian of the chain is $H=\sum_jP_j^2/(2M)+\frac{1}{2}\sum_{j,k} V_{jk}r_jr_k+H_a'$ with $P_j$ the momentum of the $j$-th atom. The force parameter $V_{jk}$ is obtained by expanding the potential energy between the $j$-th and the $k$-th atoms around their equilibrium positions $R_j$ and $R_k$ to the second order. The second term $(1/2)\sum_{j,k} V_{jk}r_jr_k$ in the Hamiltonian $H$ is for the harmonic potential. The third term $H_a'$ is for the anharmonic potential. We note that $H_a'$ is general and not limited to any particular form. We split the harmonic term into two parts by $\sum_{j,k} V_{jk}r_jr_k=\sum_{j} V_{jj}r_jr_j+\sum_{j\neq k} V_{jk}r_jr_k$ and rewrite the Hamiltonian $H$ as
\begin{align}
\label{H}
H=\sum_j\frac{P_j^2}{2M}+\frac{1}{2}\sum_{j} V_{jj}r_jr_j+H_a
\end{align}  
by absorbing the terms of $\sum_{j\neq k} V_{jk}r_jr_k$ and $H_a'$ in $H_a$. Note that the Hamiltonian $H$ in Eq.(\ref{H})  is for the atomic chain only, and does not include the baths. In the following, we will define the local Bosons for the atomic chain and introduce weight factors to average the numbers of the local Bosons in the non-equilibrium state.\\

\subsection{Local Bosons}
We express $V_{jj}=M\omega_j^2$ with $\omega_j$ the vibration frequency of the $j$-th atom. Since all the atoms in the atomic chain are identical to each other, the frequencies $\omega_j$ should be the same for all the atoms. We simplify the notation $\omega_j$ by $\omega$. The physical meaning of the frequency $\omega$ can be understood by the view of the atoms as local oscillators bonded by springs around their own equilibrium positions. $V_{jj}$ is the force parameter of the spring of the $j$-th oscillator for the $j$-th atom and $\omega$ is the oscillation frequency of the oscillator. In such a physical picture, each individual oscillator stimulates Bosons on its own local site. Thus, the Bosons are local and denoted by local Bosons(LBs). The LBs are distinguished from the Phonons. The latter represents the collective motions of all the atoms in the chain.\\

We introduce the creation and annihilation operators
\begin{align}
\label{aoperator}
a_j^{\dagger}=\sqrt{\frac{\omega M}{2\hbar }}( r_j-i\frac{P_j}{\omega M}),~~a_j=\sqrt{\frac{\omega M}{2\hbar }}( r_j+i\frac{P_j}{\omega M})
\end{align}
for the LBs of the $j$-th atom with $i$ the imaginary unit. The displacement $r_j$ and the momentum $P_j$ expressed as the function of $a_j$ and $a_j^{\dagger}$ are substituted into Eq.(\ref{H}) to get
\begin{align}
\label{HB}
H=\sum_j (a^{\dagger}_ja_j+1/2)\hbar \omega+H_a.
\end{align}
Here, $H_a$ in Eq.(\ref{HB}) is functional of only the displacements $r$, and not of the momentum $P$. We express $r_j=(a_j^{\dagger}+a_j)\sqrt{\hbar/(2\omega M)}$ and denote $A_j=a^{\dagger}_j+a_j$ for simplicity. Thus, $H_a$ is functional of $A_j$. The term $H_a$ induces the interaction between the LBs at various atomic sites, and drives the LBs to move from atom to atom in the atomic chain.\\

\subsection{Weight factor $\beta$}
We attach the two ends of the atomic chain to two thermal baths. The two baths are at their own equilibrium states with two different temperatures. The temperature of the bath connected to the $1$-st atom is denoted by $T_H$ and the temperature of the bath connected to the $N$-th atom is by $T_L$. We set $T_H>T_L$ to apply a temperature drop on the atomic chain. The number $n$ of the LBs stimulated by the $1$-st atom is weighted by the factor of $e^{-n\hbar \omega \beta_{H}}$ while the number of LBs stimulated by the $N$-th atom is weighted by $e^{-n\hbar \omega \beta_{L}}$. The two weight factors $\beta_H=1/(k_B T_H)$ and $\beta_L=1/(k_B T_L)$ are the inverse temperatures of the two baths respectively with $k_B$ the Boltzmann constant. \\

The temperature drop applied on the atomic chain drives a thermal current from the bath of $T_H$ to the bath of $T_L$ through the chain. In the steady state, the thermal current injecting to the $j$-th atom must equal the thermal current leaving the $j$-th atom. Thus, the thermal-averaged number of the LBs at the $j$-th atom is kept to be a constant in the steady state. This statement can be generalized to all the atoms of the chain, meaning that each atomic site has its own constant number of LBs in the steady state. In order to describe the constant number of the LBs at each atomic site, we introduce weight factors for every atom to average the numbers of the LBs. We denote the weight factor by $\beta$ and will complete $\beta$ with subscripts for various physical meanings. The weight factor $\beta$ plays its role like $\beta_H$ and $\beta_L$ in averaging the numbers of the LBs. Since there is no definition of temperature in the non-equilibrium states, $\beta$ is not the meaning of inverse temperature. The details of the introduction of $\beta$ will be specified in the following. \\

\subsection{Non-equilibrium ensembles}
\label{NES}
In order to describe the weight factor $\beta$ for the steady state, we take two adjacent atoms from the atomic chain as an example. The two atoms denoted by $A$ and $B$ respectively compose a two-atom system (TAS). According to Eq.(\ref{HB}), the Hamiltonian of the TAS reads
\begin{align}
\label{eh}
H_{TAS}=H_A+H_B+H_{AB}
\end{align}
with $H_A=(a_{\scriptscriptstyle A}^{\dagger}a_{\scriptscriptstyle A}+1/2)\hbar \omega$, $H_B=(a_{\scriptscriptstyle B}^{\dagger}a_{\scriptscriptstyle B}+1/2)\hbar \omega$. $H_{AB}$ is the term $H_a$ in Eq.(\ref{HB}) for the TAS. We diagonalize the Hamiltonian $H_{TAS}$ to get the energy $E_n$ and the wave function $|\phi_n\hspace{-1mm}>$ for the $n$-th Eigen state. For the TAS quantum system, we can choose $|p_{\scriptscriptstyle A},p_{\scriptscriptstyle B}\hspace{-1mm}>$ as the basis with $p_{\scriptscriptstyle A}$ the number of LBs at the site of atom A and $p_{\scriptscriptstyle B}$ the number of LBs at atom B. The numbers $p_{\scriptscriptstyle A}$ and $p_{\scriptscriptstyle B}$ of the LBs take the values from $0$ to $+\infty$. The wave function $|\phi_n\hspace{-1mm}>$ can be expressed by the basis as $|\phi_n>=\sum_{p_{\scriptscriptstyle A},p_{\scriptscriptstyle B}}A_n^{(p_{\scriptscriptstyle A},p_{\scriptscriptstyle B})}|p_{\scriptscriptstyle A},p_{\scriptscriptstyle B}>$ with $A_n^{(p_{\scriptscriptstyle A},p_{\scriptscriptstyle B})}$ the coefficients.  \\

Considering the TAS connects their adjacent atoms in the chain, the TAS is in an non-equilibrium state when the temperature drop of $\Delta T=T_H-T_L$ applied on the whole chain. We introduce an non-equilibrium Hamiltonian(NEH)
\begin{align}
\label{neh}
\tilde{H}=\beta_AH_A+\beta_B H_B+\beta_{AB}H_{AB}
\end{align}
for the non-equilibrium ensemble of the TAS. Here, the weight factors $\beta_A$, $\beta_B$ and $\beta_{AB}$ have the unit of the inverse temperature $\beta_H$ or $\beta_L$. Thus, the NEH $\tilde{H}$ is dimensionless. For a given steady state of the whole chain, the weight factors are fixed. Generally, the weight factors $\beta_A$,$ \beta_B$ and $\beta_{AB}$ should be different from each other in the steady state. In the extreme case of an equilibrium state with $\Delta T=0$, say that the TAS is at a temperature $T$, the weight factors should be reduced to be the inverse temperature $\beta_A=\beta_B=\beta_{AB}=\beta=1/(k_BT)$. In this case, the TAS stays at a quantum state with the probability proportional to $e^{-\beta(H_A+H_B+H_{AB})}$ according to the theory of the quantum statistics. By using Eq.(\ref{neh}), we write $e^{-\beta(H_A+H_B+H_{AB})}=e^{-\tilde{H}}$ for the equilibrium state. That means the Eigen values of the NEH $\tilde{H}$ can be used for the probability of the quantum states of the TAS at the equilibrium state. Such a statement can be generalized to the non-equilibrium state of the TAS. In this study, we focus on only the steady state.\\

We will solve the weight factors in a self-consistent way by meeting the physical requirements for the steady state. Now, suppose the weight factors have been figured out. The NEH $\tilde{H}$ then has a precise expression. By diagonalizing the NEH $\tilde{H}$, we get the Eigen values $\tilde{E}_n$ and the wave function $|\psi_n>$ for the $n$-th state. The set of the wave functions $|\psi_n>$ reflects the full information of the NEH $\tilde{H}$, and can be used to describe the steady state of the TAS in the chain. We further express $|\psi_n>=\sum_m C_{mn}|\phi_m>$ with $|\phi_m>$ the set of wave functions of $H_{TAS}$ and $C_{mn}$ the coefficients. For the steady state, the ensemble of the TAS can be made through the density operator
\begin{align}
\label{rho}
\hat{\rho}=\frac{1}{Z}\sum_n e^{-\tilde{E}_n}|\psi_n><\psi_n|
\end{align} 
with $Z$ the partition function equaling $Z=\sum_n e^{-\tilde{E}_n}$.\\

If we prepare a steady state for the atomic chain and the TAS is stabilized at the $n$-th state with the wave function of $|\psi_n>$, then we decouple the TAS from the atomic chain and isolate the TAS. The quantum state of the isolated TAS(ITAS) will evolve with time, which is denoted by $|\psi_n(t)>$. The initial state $|\psi_n(0)>$ is exactly $|\psi_n>$. We emphasize that the time evolving of $|\psi_n(t)>$ is not governed by the NEH $\tilde{H}$ since $\tilde{H}$ is not the Hamiltonian of the ITAS. Instead, it evolves with the Hamiltonian $H_{TAS}$ by
\begin{align}
\label{psint}
|\psi_n(t)>&=\sum_mC_{mn}e^{iE_mt/\hbar}|\phi_m>\nonumber\\
&=\sum_m^{p_{\scriptscriptstyle A},p_{\scriptscriptstyle B}}C_{mn}A_m^{(p_{\scriptscriptstyle A},p_{\scriptscriptstyle B})}e^{iE_mt/\hbar}|p_{\scriptscriptstyle A},p_{\scriptscriptstyle B}>.
\end{align}
In the second line of the above expression, we have used $|p_{\scriptscriptstyle A},p_{\scriptscriptstyle B}>$ as the basis. Then, in the time evolving of the ITAS with the initial state $|\psi_n>$, the probability of the ITAS at the state of $|p_{\scriptscriptstyle A},p_{\scriptscriptstyle B}>$ is
\begin{align}
\mathcal{P}(n,p_A,p_B,t)=\sum_{m,q}D(n,m,q,p_{\scriptscriptstyle A},p_{\scriptscriptstyle B})e^{i(E_q-E_m)t/\hbar}
\end{align}
with $D(n,m,q,p_{\scriptscriptstyle A},p_{\scriptscriptstyle B})=[C_{mn}A_m^{(p_A,p_B)}]^{\dagger}C_{qn}A_q^{(p_A,p_B)}$. The density operator for the ITAS after the decoupling of the TAS from the atomic chain is time dependent, reading
\begin{align}
\hat{\rho}(t)=\frac{1}{Z}\sum_n e^{-\tilde{E}_n}|\psi_n(t)><\psi_n(t)|.
\end{align} 

\subsection{Thermal current}
In the ITAS, the averaged numbers of LBs at atom A and atom B both are time dependent, and denoted by  $\bar{p}_A(t)$ and  $\bar{p}_B(t)$ respectively. We have $\bar{p}_A(t)=Tr[a_A^{\dagger}a_A\hat{\rho}(t)]$ and $\bar{p}_B(t)=Tr[a_B^{\dagger}a_B\hat{\rho}(t)]$ with \textit{Tr} representing the traces of matrices. Explicitly, they are 
\begin{align}
\label{n1t}
&\bar{p}_A(t)=\sum_{n}^{p_A,p_B}p_A \frac{e^{-\tilde{E}_n}}{Z}\mathcal{P}(n,p_A,p_B,t),\\
\label{n2t}
&\bar{p}_B(t)=\sum_{n}^{p_A,p_B}p_B \frac{e^{-\tilde{E}_n}}{Z}\mathcal{P}(n,p_A,p_B,t).
\end{align}  
Here, the initial values  $\bar{p}_A(0)$ and $\bar{p}_B(0)$ are the averaged numbers of the LBs for the TAS when the TAS is still connected in the atomic chain at the steady state before the decoupling.\\

The total energy of the ITAS is conserved and LBs will flow in the ITAS, such as from atom A to atom B or reverse.  The flowing of the LBs in the ITAS behaves like the Rabi oscillation, which will be specified in the Sec.(\ref{cal}) later. Here, we use the changing of $\bar{p}_A(t)$ and $\bar{p}_B(t)$ to define the thermal current. After the time duration $\Delta t$, the number of the LBs at atom A will change from $\bar{p}_A(0)$ to $\bar{p}_A(\Delta t)$ . Thus, the thermal current flowing to atom A can be defined by 
\begin{align}
\label{JA}
J_A=\hbar \omega\frac{\bar{p}_A(\Delta t)-\bar{p}_A(0)}{\Delta t}
\end{align}
in average. Similarly, we define the thermal current flowing to atom B by 
\begin{align}
\label{JB}
J_B=\hbar \omega \frac{\bar{p}_{_B}(\Delta t)-\bar{p}_{_B}(0)}{\Delta t}
\end{align}
averaged in the time duration $\Delta t$. The thermal current $J_A$ flowing to atom A can be effectively considered as the thermal current $-J_A$ leaving atom A to atom B. The thermal current $-J_A$ normally is different to the thermal current $J_B$ flowing to atom B. That means after the time duration $\Delta t$ there is an amount of thermal energy $(-J_A-J_B)\Delta t$ absorbed by the interaction energy $H_{AB}$. Thus, the interaction energy $H_{AB}$ may exchange energy with $H_A+H_B$, and make the thermal currents $J_A$ and $J_B$ different in absolute value. \\

The thermal currents $J_A$ and $J_B$ flowing between atom A and atom B can be considered to relax the ITAS from its steady state just after decoupled from the atomic chain. Such a relaxation process also occurs for the TAS when it is still connected in the chain if we remove the temperature drop from the atomic chain. In order to keep the steady state of the atomic chain and keep the non-equilibrium state of the TAS, the relaxation process of the TAS should be blocked by exchanging energy between the TAS and the remaining part of the chain. In the atomic chain, there are two atoms adjacent to atom A. One is atom B in the TAS and the other is denoted by $A'$ out of the TAS. Similarly, we denote the atom adjacent to atom B by $B'$ out of the TAS. That means, in the steady state, a thermal current with the value of $J_A$ is flowing from atom $A'$ to atom A, to compensate for the thermal current $-J_A$ leaving atom A to atom B. Only in this way, $\bar{p}_A(0)$ is kept to be a constant for atom A in the steady state. Similarly, a thermal current $-J_B$ needs to flow from atom B to atom $B'$ to compensate for the thermal current $J_B$ flowing from atom A to atom B in the TAS, by which $\bar{p}_B(0)$ can be kept as a constant in the steady state. Therefore, the thermal current $J_A$ flowing from the atom $A'$ to the atom A and the thermal current $-J_B$ leaving atom $B$ to atom $B'$ require that an equation
\begin{align}
\label{jzero}
J_A=-J_B
\end{align}
must be held for the steady state.\\

We have defined the thermal currents in Eq.(\ref{JA}) and Eq.(\ref{JB}). A question remaining is how to choose the time duration $\Delta t$ for the calculation. In the Sec(\ref{cal}), we will show that $\Delta t$ should be chosen to maximize the absolute value of  the thermal current. In this way, the ITAS can relax itself from the non-equilibrium state as soon as possible~\cite{Prig71}.\\

\subsection{Algorithm}
\label{algorithm}
For the TAS in the atomic chain at the steady state, totally we have defined seven variables. Three weight factors $\beta_A, \beta_B, \beta_{AB}$, two averaged numbers $\bar{p}_A(0), \bar{p}_B(0)$ of the LBs, and two thermal currents $J_A, J_B$. And we have five equations  Eq.(\ref{n1t},\ref{n2t},\ref{JA},\ref{JB},\ref{jzero})in hand. In practice, if $\beta_A, \beta_B$ and $\beta_{AB}$ are given, $\bar{p}_A(0)$ and $\bar{p}_B(0)$ can be solved from Eq.(\ref{n1t},\ref{n2t}) while $J_A$ and $J_B$ can be solved from Eq.(\ref{JA},\ref{JB}). In the later calculations, we find that Eq.(\ref{jzero}) is equivalent to the condition of  $\beta_{AB}=(\beta_A+\beta_B)/2$. Thus, we need only other two conditions to determine $\beta_A$ and $\beta_B$. Actually, the two conditions can be obtained by the fixed temperatures of the two baths connected to the atomic chain.\\

To solve the thermal conductivity of the whole atomic chain, we present the algorithm in the following.
\begin{enumerate}
\item{\label{stp1}
Starting from the $N$-th atom and setting $\beta_N=1/(k_B T_L)$ for the atom.}
\item{\label{stp2} Assuming a positive thermal current $J$ for the atomic chain. The direction of $J$ is from the bath of $T_H$ to the bath of $T_L$.}

\item{\label{stp3}
Choosing a value $\beta_{N-1}$ less than $\beta_N$ for the $(N-1)$-th atom. 
\begin{itemize}
\item{Set the weight factor $\beta_{(N-1)N}=(\beta_N+\beta_{N-1})/2$ for the interaction energy between the $N$-th and the $(N-1)$-th atoms. }
\item{Complete the NEH $\tilde{H}$ of Eq.(\ref{neh}) for the TAS comprising the $(N-1)$-th and the $N$-th two atoms with the weight factors $\beta_{N-1},\beta_N$ and $\beta_{(N-1)N}$. }
\item{Assign the $(N-1)$-th atom by atom A and the $N$-th atom by atom B. Get $\bar{p}_{N-1}(t)$ through Eq.(\ref{n1t}) and solve the thermal current $J_A=-J'$ from Eq.(\ref{JA}). If $J'\neq J$, then change the value $\beta_{N-1}$ and repeat the step 3 until getting a proper value $\beta_{N-1}$ to meet the requirement $J'=J$ within the accuracy.}
\end{itemize}
}
\item{\label{stp4} Choosing a value $\beta_{j}$ for the $j$-th atom with $\beta_{j+1}$ determined from the last step already. Following the list in the step \ref{stp3} with the index $(N-1)$ replaced by the index $j$ and $N$ by $j+1$. Finally, get the proper value $\beta_{j}$ to get the thermal current equaling $J$.}
\item{\label{stp5}Running the index $j$ from $N-2$ to $1$ by following the step \ref{stp4}. Finally, get $\beta_1$ for the $1$-st atom. If $\beta_1\neq 1/(k_B T_H)$, then return to step \ref{stp2} and assume a new thermal current $J$ for the atomic chain. After that, repeat the steps \ref{stp3},\ref{stp4} and \ref{stp5}, until a proper thermal current $J$ is obtained by which $\beta_1= 1/(k_B T_H)$ is satisfied.}
\item{Calculating the thermal conductivity by using data of the temperature drop $T_H-T_L$, the thermal current $J$ and the length $L$ of the atomic chain.}
\end{enumerate}
In the above algorithm, $\beta_1$ and $\beta_N$ are fixed for the atomic chain and the quantum dynamics have been implemented for the transport of the LBs. All the weight factors and the thermal currents are solved in a self-consistent way.\\
 
\section{calculations}
\label{cal}
To reach the thermal conductivity of the atomic chain, we need to calculate the quantum state of the ITAS to define the thermal currents. The ITAS provides basic knowledge for the study. Therefore, we need an explicit form for the term $H_{AB}$ in Eq.(\ref{eh}). Practically, we write the interaction energy $H_{AB}$ for the TAS as
\begin{align}
H_{AB}&=U_{2}A_AA_B+\frac{1}{3!}\sum_{q,j,k}U_{qjk}A_qA_jA_k\nonumber\\
&+\frac{1}{4!}\sum_{q,j,k,l}U_{qjkl}A_qA_jA_kA_l,
\end{align}
in which the first term in the right hand side is from the harmonic potential and the last two terms are for the anharmonic potentials. The subscripts $q,j,k,l$ run over the two atoms $A$ and $B$. We have defined $A_q=a_q+a_q^{\dagger}$ before. The coefficients $U_{qjk}$ and $U_{qjkl}$ can be modified to enhance the anharmonic interaction of the TAS. The relations between the coefficients are simplified to be $U_{AAA}=U_{ABB}=-U_3$, $U_{AAB}=U_{BBB}=U_3$, $U_{AAAA}=U_{AABB}=U_{BBBB}=U_4$ and $U_{AAAB}=U_{ABBB}=-U_4$ for the calculations.\\

For clarity, we normalize the temperature $T$ by $T_0=100K$, the frequency $\omega$ by $\omega_0=k_BT_0/\hbar=1.31\times 10^{13}Hz$, and the energy scale by $\hbar \omega_0$. By using the parameters of the Silicon atom, we normalize the length scale by $l_0=0.21nm$, the atomic mass by $M_0=4.65\times 10^{-26}kg$, time scale by $t_0=l_0\sqrt{M_0/(k_B T_0)}=1.22\times 10^{-12}s$. The thermal current is normalized by $J_0=\hbar \omega_0 /t_0=1.13\times 10^{-9}Jol/s$. Here, we use $Jol$ to represent the unit of energy, in order to distinguish from the symbol $J$ for the thermal current. The thermal conductivity is defined as $\kappa=JL/\Delta T$, which is normalized by $\kappa_0=J_0l_0/T_0=2.375\times 10^{-21}Jol\cdot m/(K \cdot s)$. \\

The parameters used in the calculations are listed as $l=1.12$, $M=1$ and $\omega=0.6$. The harmonic coefficient $U_2$ is set to be $U_2=-\omega/4$, equaling $U_{2}=-0.15$. The details of $U_2$ could be found in Appendix \ref{appA}. The anharmonic coefficients $U_3$ and $U_4$ will be specified later. \\

\subsection{LB numbers $\bar{p}_A(t)$ and $\bar{p}_B(t)$ }
The numbers of the LBs for the ITAS are time dependent after the decoupling of the TAS from the atomic chain. The averaged numbers have been denoted by $\bar{p}_A(t)$ and $\bar{p}_B(t)$ for atom A and atom B respectively. $\bar{p}_A(0)$ and $\bar{p}_B(0)$ are equivalent to the averaged numbers of the LBs for the TAS at the steady state. We plot $\bar{p}_A(t)$ and $\bar{p}_B(t)$ in Fig.1. In the plot, we set $\beta_A=0.834$,  $\beta_B=1$ and vary $\beta_{AB}$. The dimension of the Boson space takes the value of 10. And the anharmonic coefficients take the values of $U_3=0.08$ and $U_4=0.008$.
\begin{figure}
\includegraphics[scale=0.25]{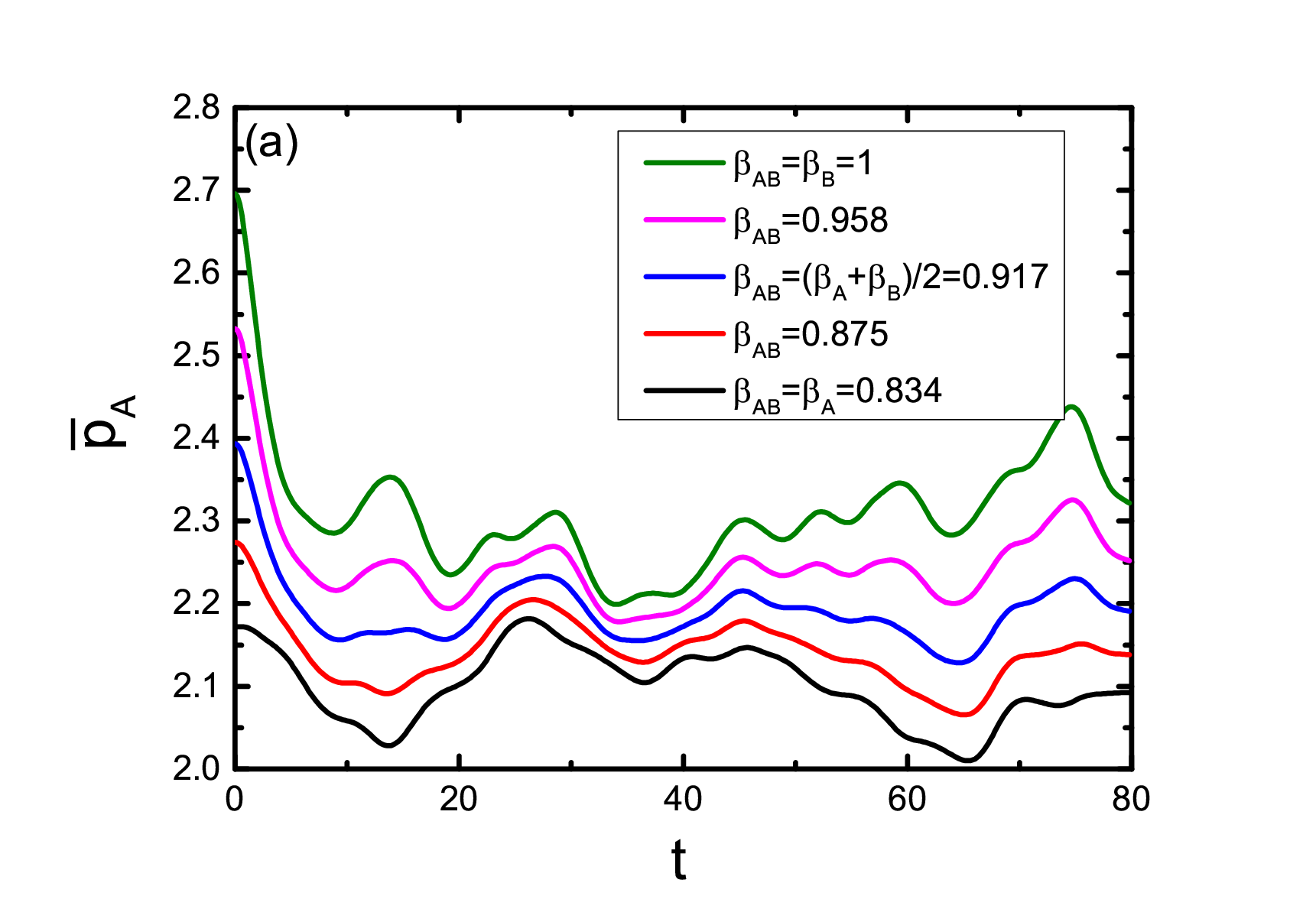}
\includegraphics[scale=0.25]{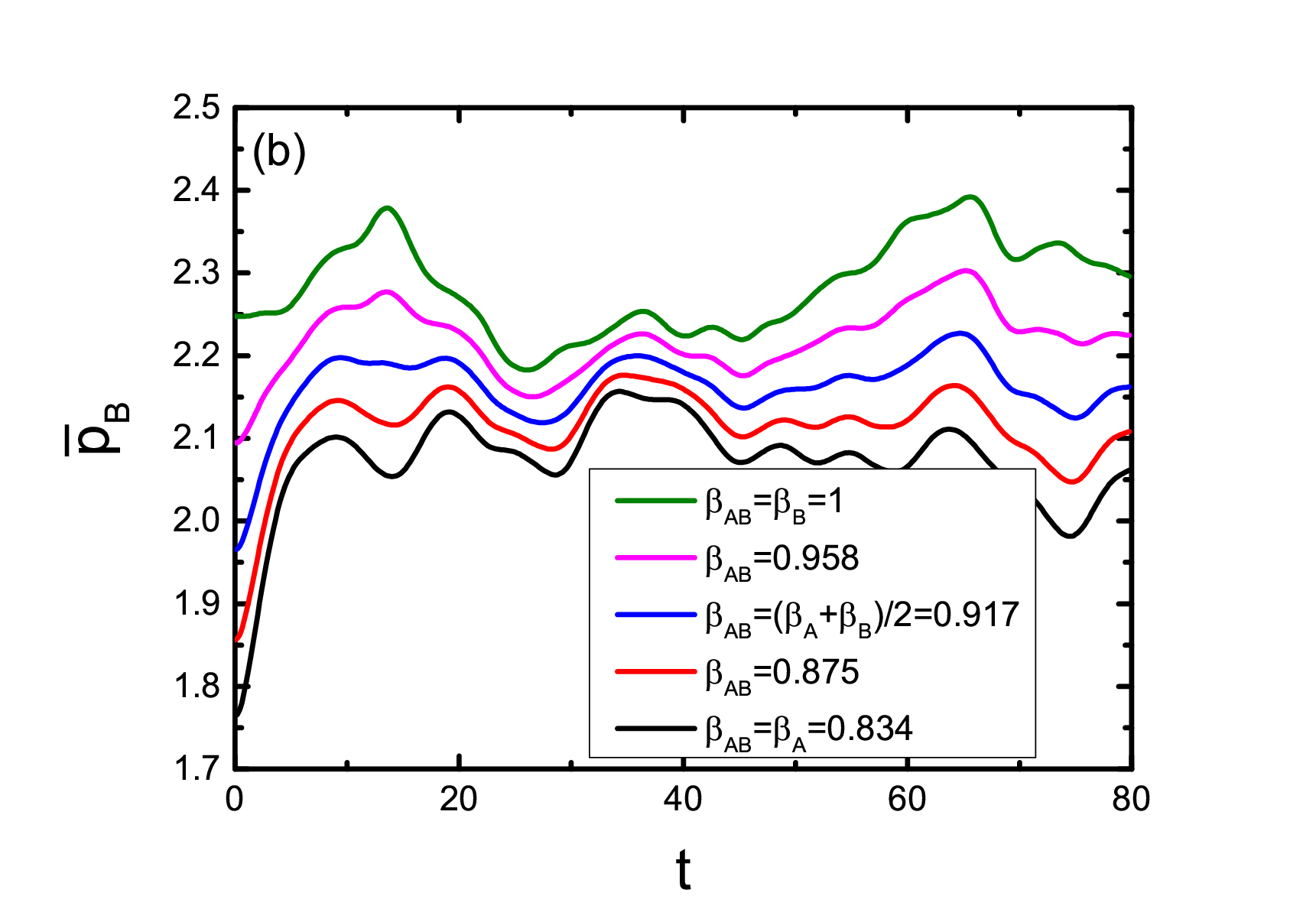}
\includegraphics[scale=0.25]{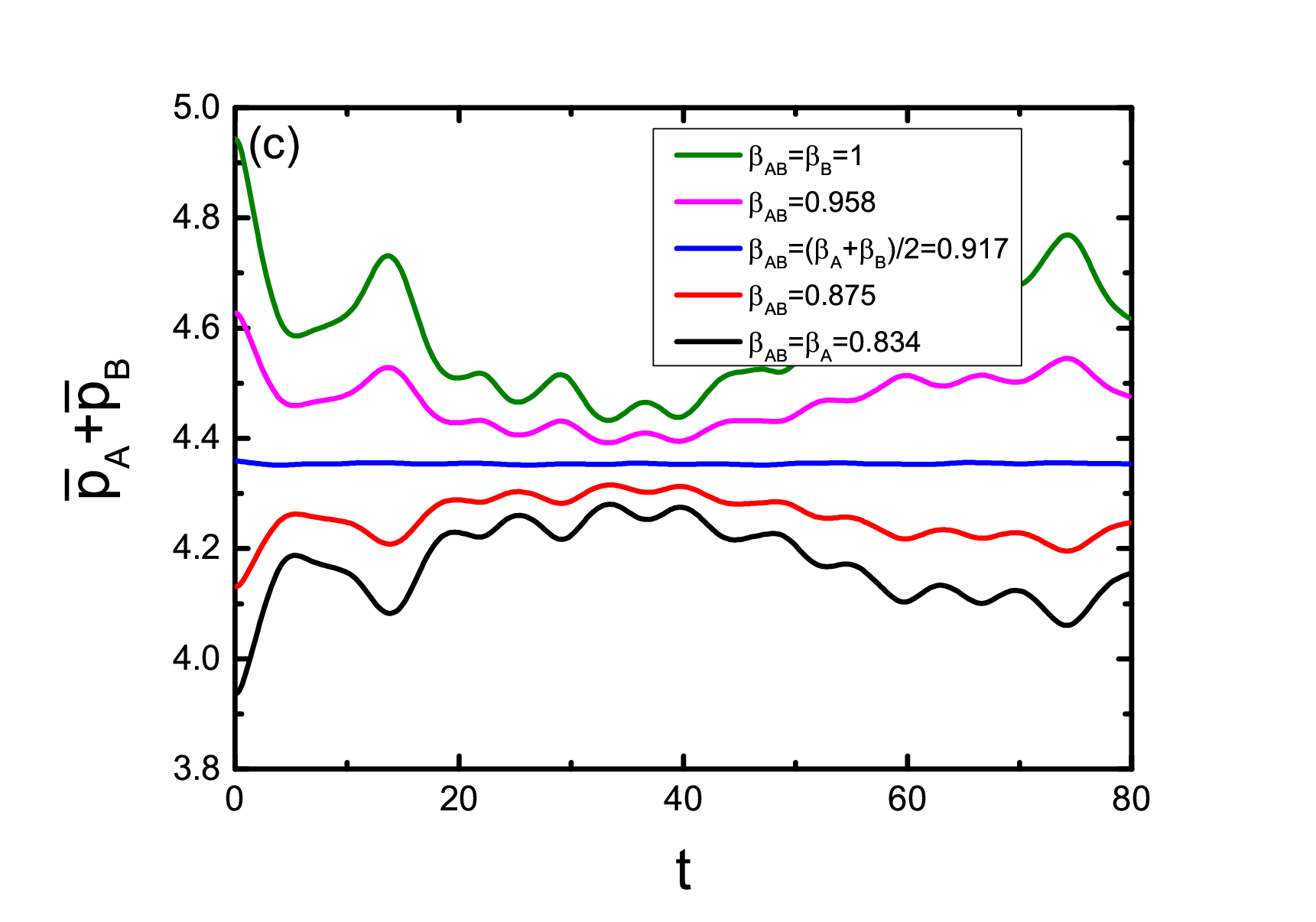}
\includegraphics[scale=0.25]{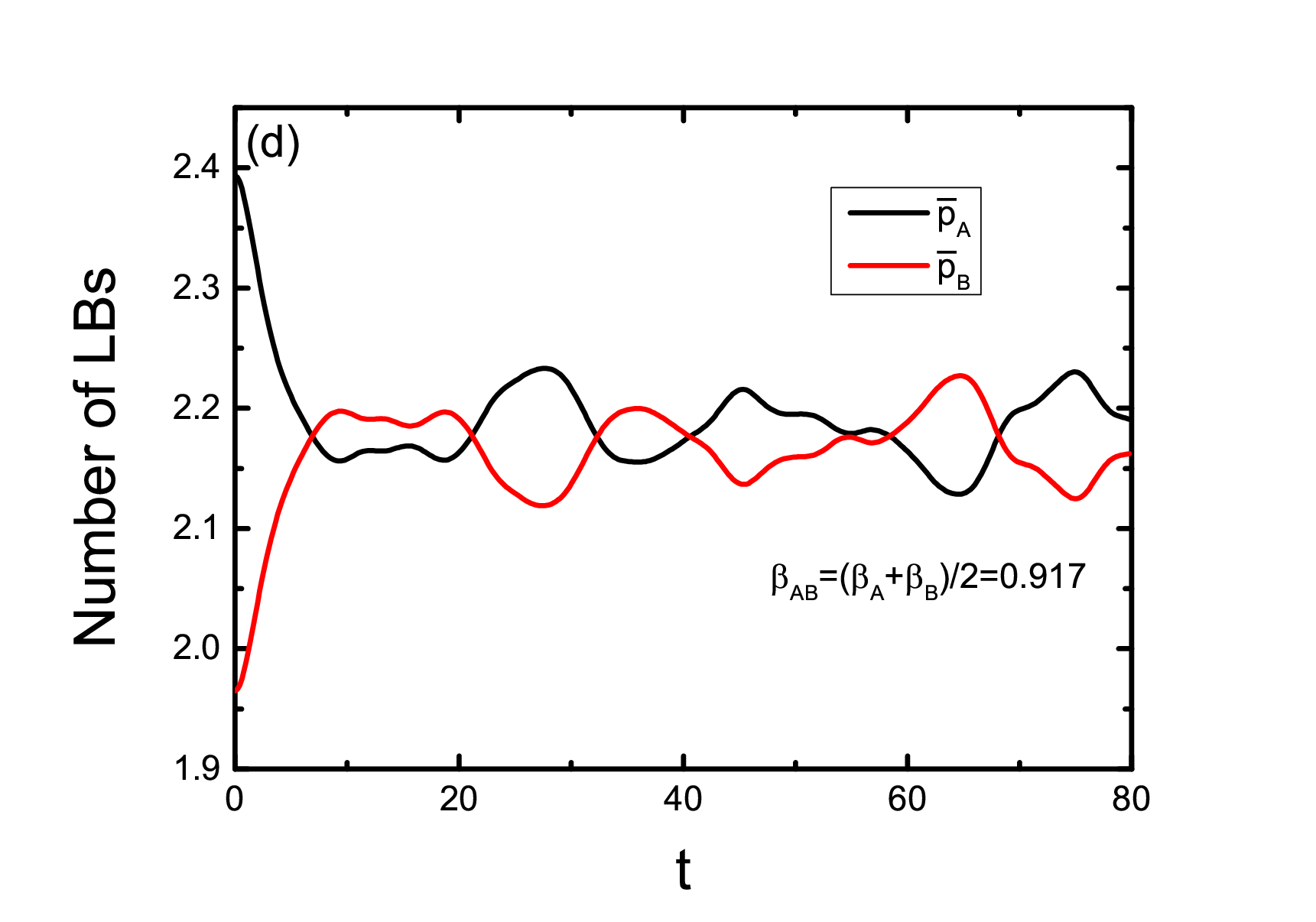}
\caption{Averaged numbers of Local Bosons(LBs) dependent on time in the ITAS.The weight factors take the values of $\beta_A=0.834$ and $\beta_B=1$. (a) $\bar{p}_A$ is the averaged number of the LBs for atom A. (b)$\bar{p}_B$ is the averaged number of the LBs for atom B. (c) $\bar{p}_A+\bar{p}_B$ is the total averaged number of the LBs in the ITAS.(d)The time dependence of $\bar{p}_A$ and $\bar{p}_B$ behaves like Rabi oscillation with $\beta_{AB}=(\beta_A+\beta_B)/2=0.917$ fixed.}
\end{figure}
\\

It could be found in Fig.1(a) that the function $\bar{p}_A(t)$ is oscillating with the time. Such oscillating phenomena of $\bar{p}_A(t)$ can be understood from the Rabi oscillation. It has been mentioned in Sec.(\ref{NES}) that the $n$-th Eigen state of the ITAS is denoted by $|\phi_n>$ with the Eigen energy $E_n$. The wave function $|\phi_n>$ is varying with the time by $e^{-iE_n t/\hbar}$. If the ITAS is prepared in the $n$-th Eigen state $|\phi_n>$, the number of the LBs is a constant at each site of the atoms, and is time independent. But now, the initial state of the ITAS is the thermal state $|\psi_n>$, not the Eigen state $|\phi_n>$. The thermal state $|\psi_n>$ is composed by $|\phi_n>$ as shown in Eq.(\ref{psint}). The Eigen states $|\phi_n>$ in the thermal state $|\psi_n>$ interacts with each other, leading to the LBs flowing between the eign states $|\phi_n>$ to form the Rabi-like oscillation. Similar oscillation behaviors have been found in Fig.1(b), which is the plot of $\bar{p}_B(t)$ for atom B in the ITAS.\\

The Hamiltonian $H_{TAS}$ in Eq.(\ref{eh}) can be split into two parts. One part is $H_A+H_B$ and the other part is $H_{AB}$. The thermal energy may transfer between the two parts. We denote the total number of the LBs in the ITAS by $\mathcal{N}(t)=\bar{p}_A(t)+\bar{p}_B(t)$. The total energy of $H_A+H_B$ in Eq.(\ref{eh}) is $\hbar \omega \mathcal{N}(t)$. The potential $H_{AB}$ may absorb energy from $H_A+H_B$ to decrease $\mathcal{N}$, or release energy to $H_A+H_B$ to increase $\mathcal{N}$. Thus, the total numbers $\mathcal{N}(t)$ of the LBs can be expected to oscillate with time. We plot $\mathcal{N}(t)$ in Fig.1(c), and find the oscillation behaviors expected for $\mathcal{N}(t)$. In the meantime, we find that $\mathcal{N}(t)$ can be time independent when the weight factor $\beta_{AB}$ takes the value of $\beta_{AB}=(\beta_A+\beta_B)/2=0.917$. The time independence of $\mathcal{N}(t)$ means that there is no energy transferring between $H_A+H_B$ and $H_{AB}$ if $\beta_{AB}=(\beta_A+\beta_B)/2$ is fixed. And the LBs flow between atom A and atom B directly without being absorbed or released by $H_{AB}$. In this way, Eq.(\ref{jzero}) of $J_A=-J_B$ is satisfied. Such a result has been confirmed by our numerical calculations by varying $\beta_A$ and $\beta_B$ from $0.3$ to $3$(data not shown here), but has not been understood theoretically. The theoretical understanding of the condition $\beta_{AB}=(\beta_A+\beta_B)/2$ equivalent to the condition $J_A=-J_B$ of Eq.(\ref{jzero}) is an open question. In the Sec.(\ref{algorithm}), we have adopted $\beta_{j(j+1)}=(\beta_{j}+\beta_{j+1})/2$ for the TAS comprising the $j$-th and the $(j+1)$-th atoms. For clarity, we plot the numbers of the LBs with $\beta_{AB}=(\beta_{A}+\beta_{B})/2$ fixed in Fig.1(d). In the figure, it could be found that $\bar{p}_A(0)>\bar{p}_B(0)$ which can be understood in the following. The probability of the quantum state of the TAS in the steady state is proportional to $e^{-\tilde{E}}=e^{-(\beta_A p_A+\beta_B p_B)\hbar \omega-\beta_{AB}H_a}$ as shown in Eq.(\ref{rho}). We have set $\beta_A<\beta_B$, indicating that the quantum states with $p_A>p_B$ have the higher probability than the states with $p_A\leq p_B$. With the time evolving,  $\bar{p}_A(t)$ decreases and $\bar{p}_B(t)$ increases, which can be used to define the thermal current by Eq.(\ref{JA}) and Eq.(\ref{JB}). We emphasize that a closed quantum system has time reversal symmetric dynamics. The numbers of $\bar{p}_A(0)$ and $\bar{p}_B(0)$ will be recovered after a long time duration for the ITAS. However, the recovering does not influence our definition of the thermal current. \\

\subsection{Thermal currents $J_A$ and $J_B$}
By using the data of Fig.1, we calculate  $J_A$ and $J_B$ from the definitions Eq.(\ref{JA}) and Eq.(\ref{JB}), and plot the results in Fig.2.
\begin{figure}
\includegraphics[scale=0.25]{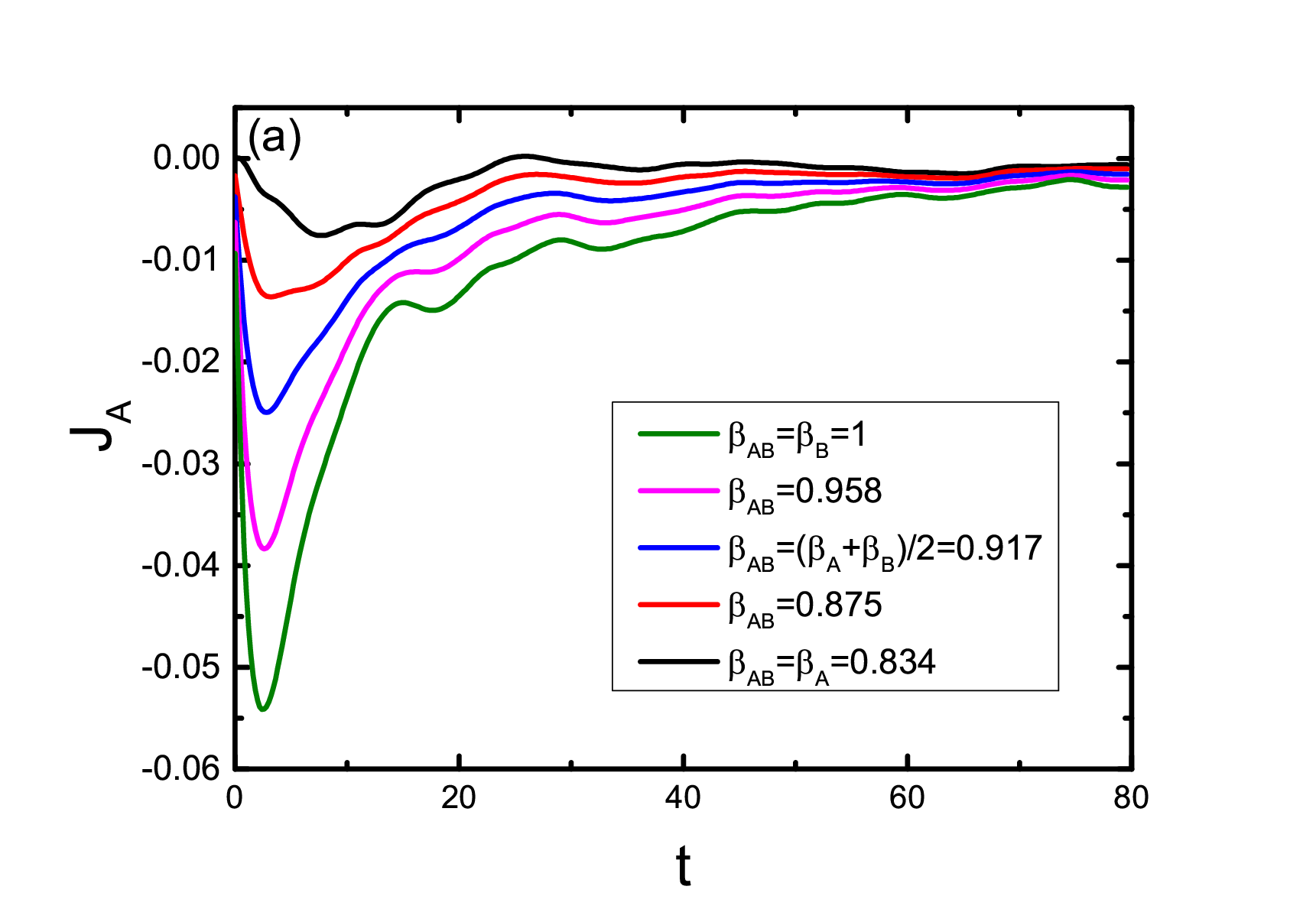}
\includegraphics[scale=0.25]{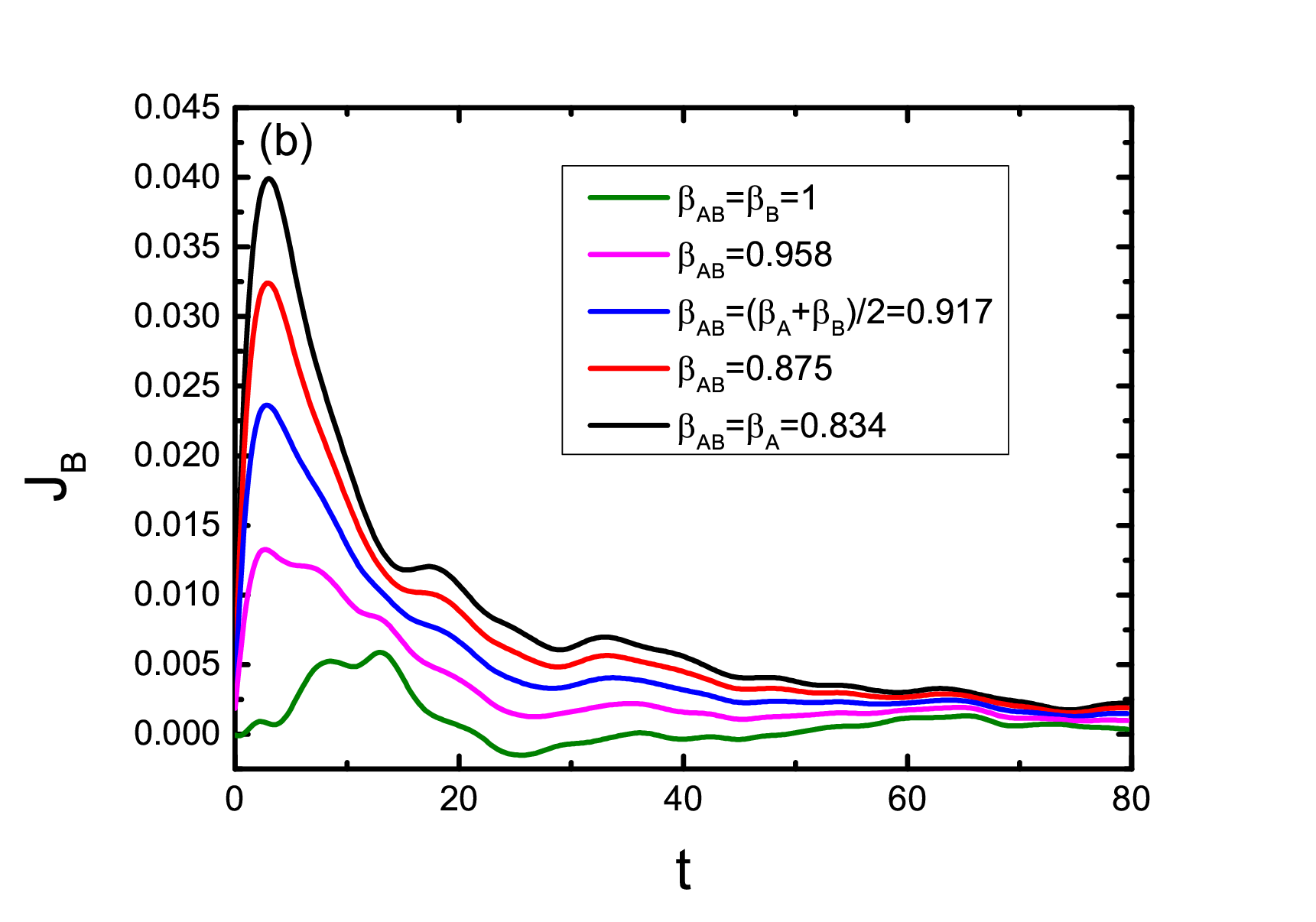}
\includegraphics[scale=0.25]{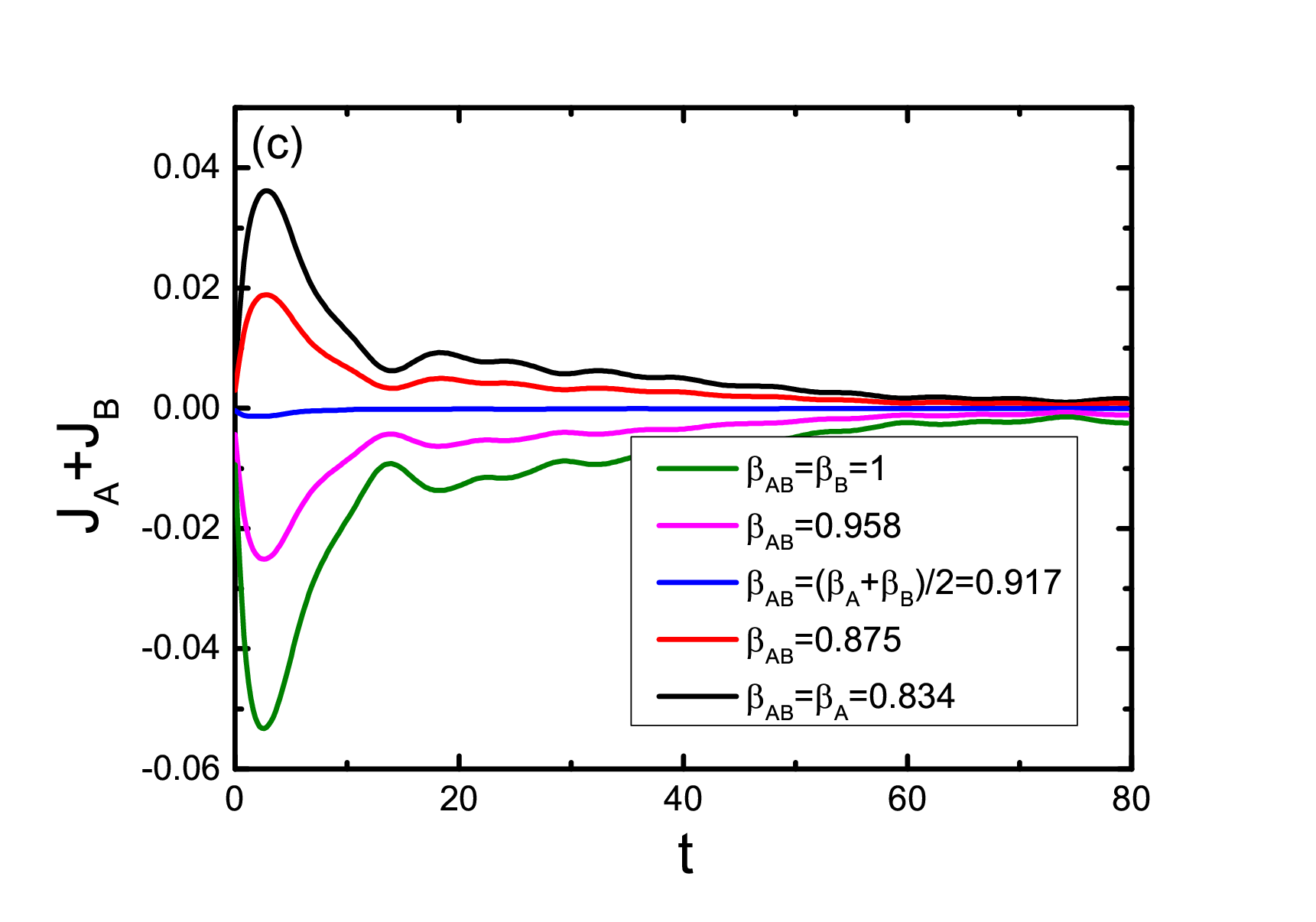}
\caption{Thermal currents dependent on time in the ITAS.(a)$J_A$ is the thermal current flowing to atom A.(b)$J_B$ is the thermal current flowing to atom B.(c)$J_A+J_B$ indicates the thermal energy absorbed or released by the interaction potential between the two atoms. Only the weight factor $\beta_{AB}=(\beta_A+\beta_B)/2$ guarantees that there are no LBs absorbed or released by the interaction potential in the ITAS.}
\end{figure} 
With the time increasing, the absolute values of the thermal currents $|J_A|$ and $|J_B|$ both increase and then decay to zero after reaching peaks, as shown in Fig.2(a) and (b) respectively. The decaying to zero is originated from the enlargement of $\Delta t$ as the denominator. In Fig.2(c), we plot $J_A+J_B$ by varying $\beta_{AB}$ and find that when $\beta_{AB}=(\beta_A+\beta_B)/2$ is held, $J_A+J_B=0$ in Eq.(\ref{jzero}) can be guaranteed. In the following study, we will fix $\beta_{AB}=(\beta_A+\beta_B)/2$ for the thermal current. And the thermal current $J_A$ or $J_B$ is determined at the maximum of its absolute value. This is because the maximum of the absolute value of the thermal current can lead to the maximum entropy production for the transient process before reaching the principle of the minimum entropy production for the steady state ~\cite{Prig71}.

\subsection{Thermal conductivity}
According to the algorithm in the Sec.(\ref{algorithm}), we calculate the weight factors of the atomic chain. $\beta_j$ is the weight factor for the $j$-th atom. The weight factors $\beta_1=1/(k_B T_H)$ and $\beta_N=1/(k_B T_L)$ are the inverse temperatures of the two baths respectively, which have been fixed. All the other weight factors $\beta_j$ are not the inverse temperatures, since there is no definition of the temperature in the non-equilibrium state. Considering that the weight factor plays its role to average the number of the LBs, it can be regarded as the inverse temperatures effectively. We can define the effective temperature as $T_j=1/(k_B \beta_j)$ for the $j$-th atom. Based on the data of the weight factors, the effective temperatures can be obtained and are plotted in Fig.3(a) for the atomic chain having $N=80$ atoms. 
\begin{figure}
\includegraphics[scale=0.25]{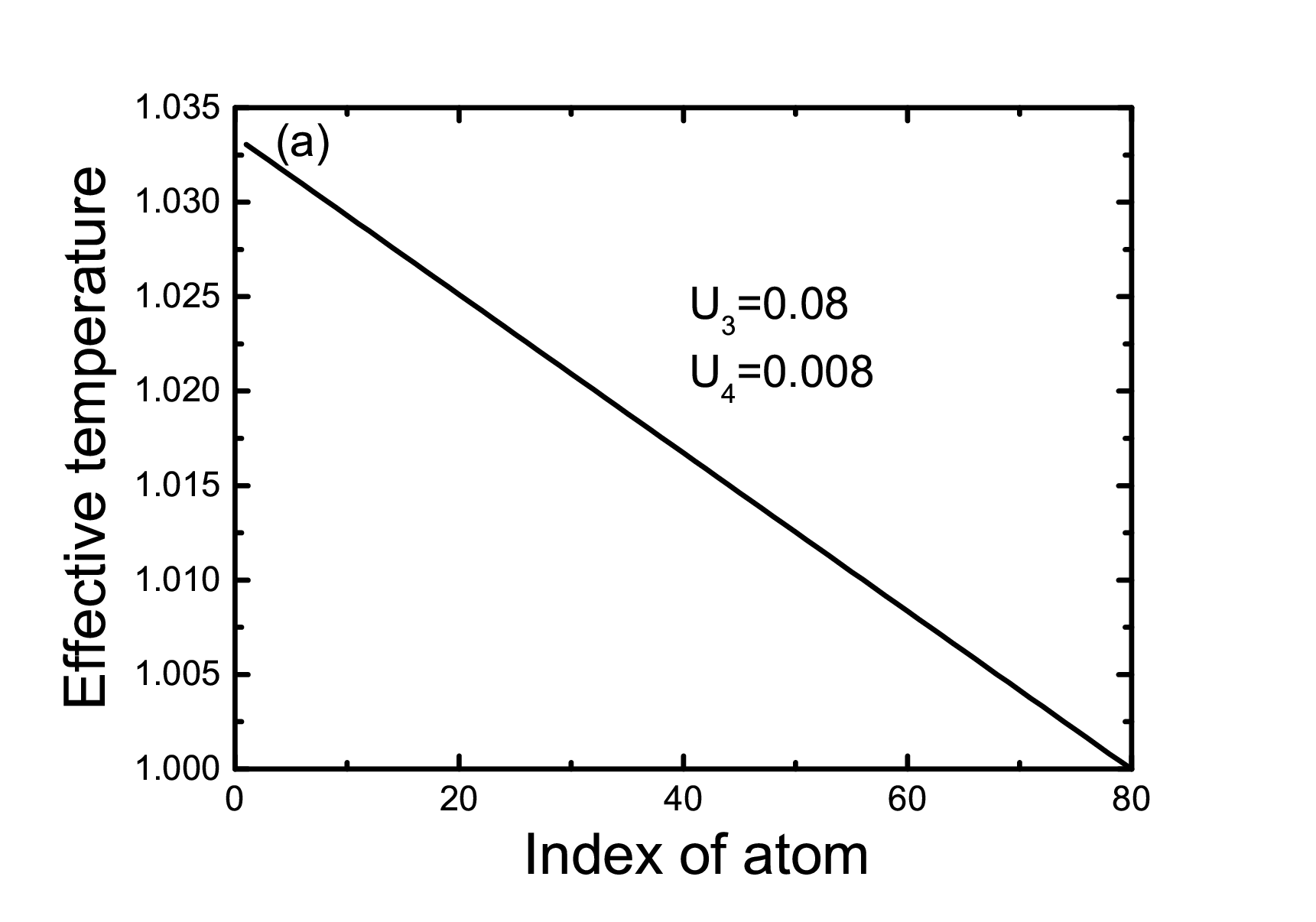}
\includegraphics[scale=0.25]{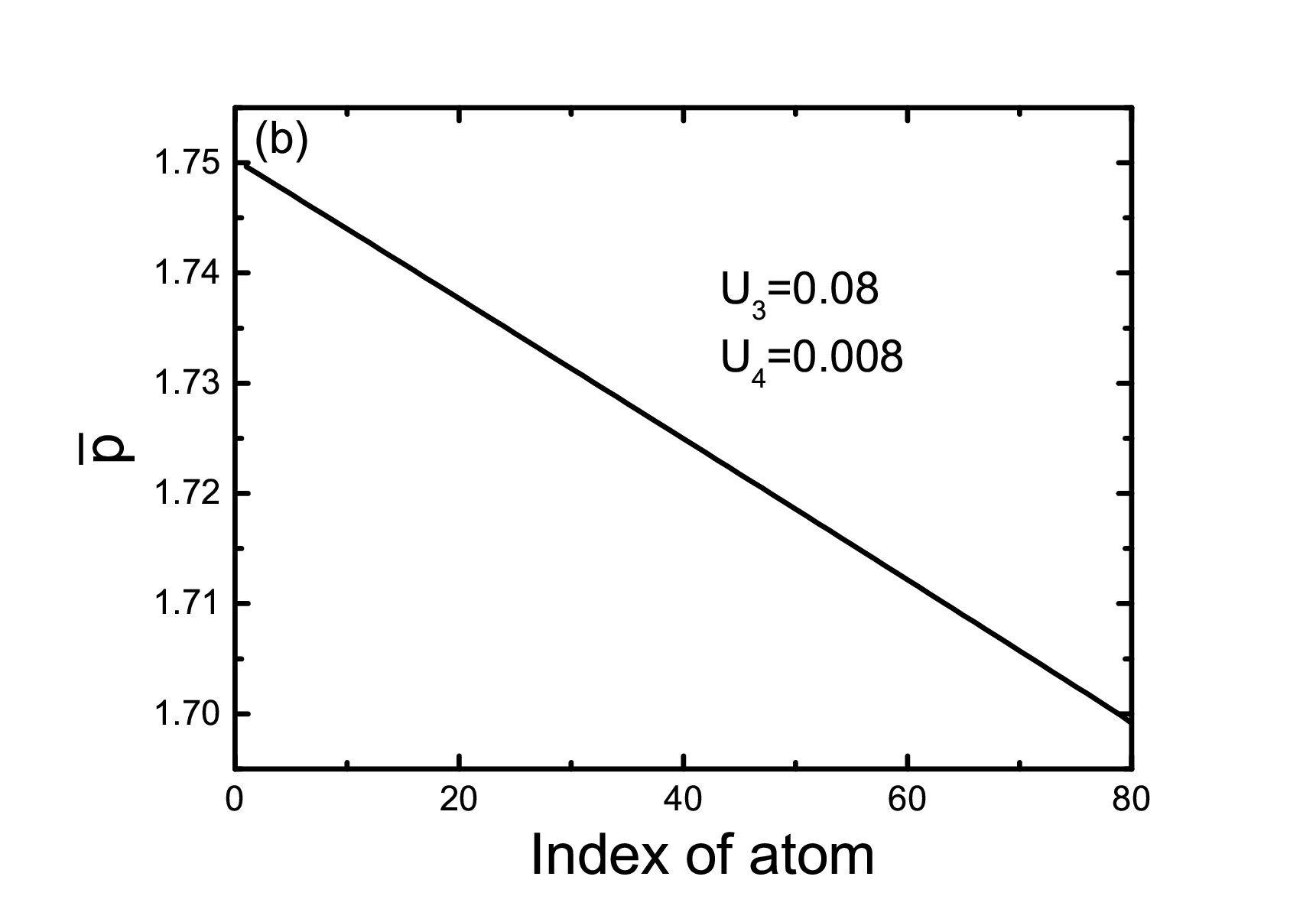}
\includegraphics[scale=0.25]{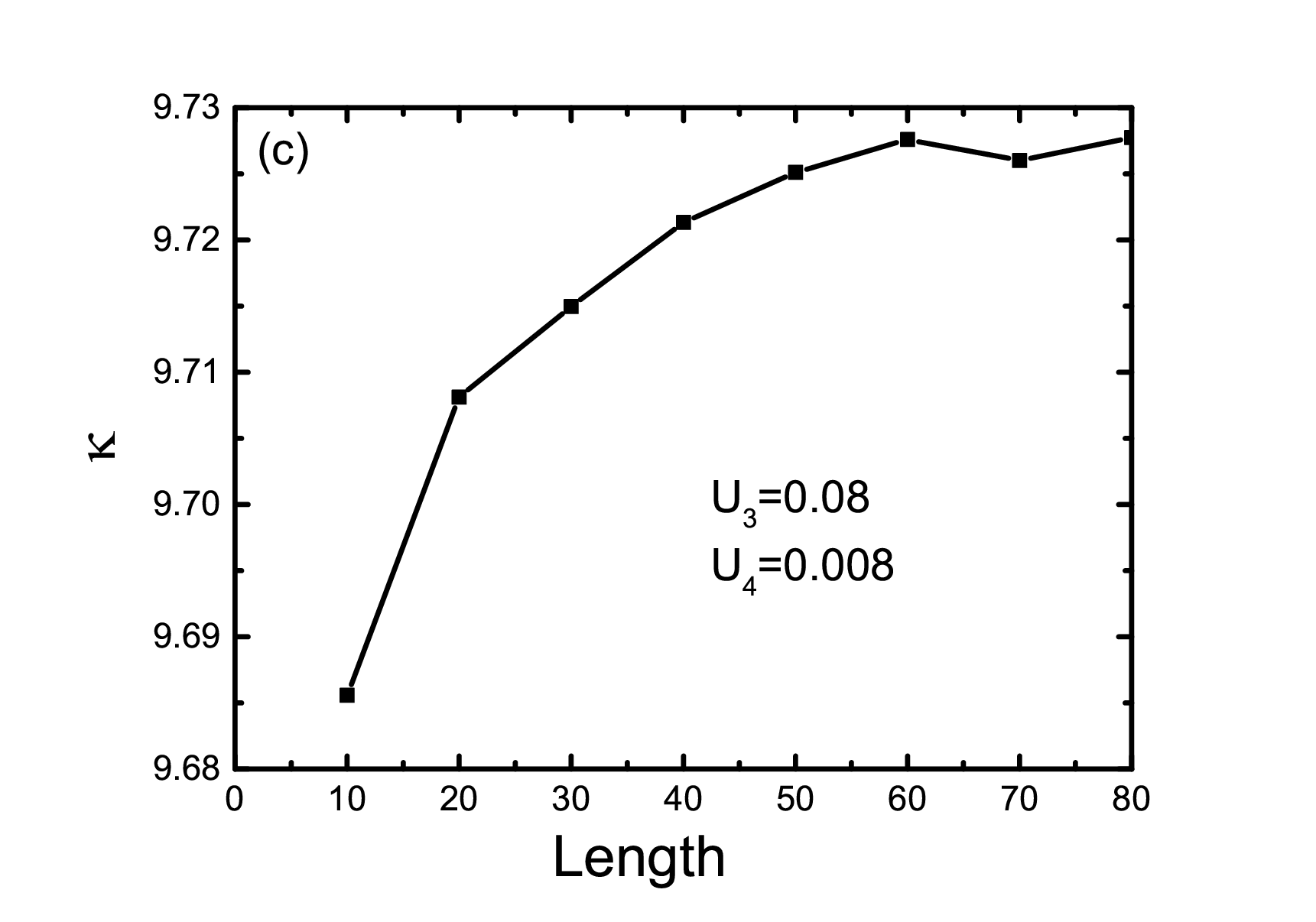}
\includegraphics[scale=0.25]{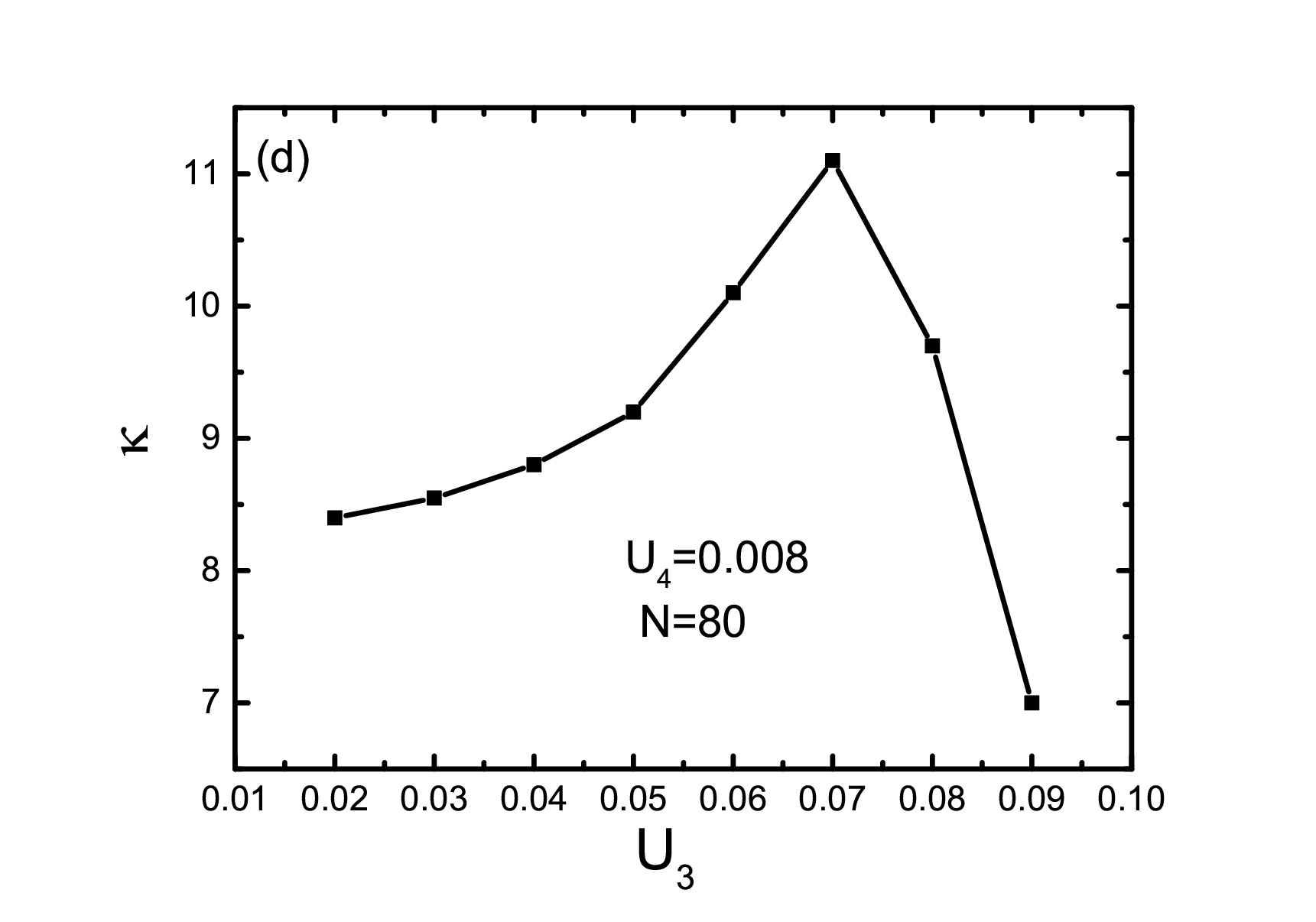}
\caption{Thermal transport in the atomic chain.(a)The effective temperature in the atomic chain is linear.(b)The averaged number $\bar{p}$ of the LBs at each atomic site shows its linearity in the atomic chain.(c)Thermal conductivity $\kappa$ varies with the length of the atomic chain. (d)Thermal conductivity $\kappa$ is dependent on the anharmonic coefficient $U_3$.}
\end{figure} 
In Fig.3(a), we set $T_H=1.033$, $T_L=1$, $U_3=0.08$ and $U_4=0.008$. The effective temperatures $T_j$ are linear in the chain. In Fig.3(b), we plot the averaged number $\bar{p}$ of the LBs at each atomic site, which is found also to be linear. \\

We have defined the thermal conductivity by $\kappa=(T_H-T_L)/(JL)$ and calculated $\kappa$ for atomic chains with various lengths. Results have been plotted in Fig.3(c), showing that $\kappa$ increases with the length of the atomic chain. It has been confirmed by experiments that the thermal conductivity $\kappa$ increases with the length of the nanostructure~\cite{Chen13}. However, our results can not be used to fit the experimental results since our model in present study is a toy model, in which the interaction potential between atoms has been simplified. And the toy model does not consider the fluctuations of the thermal current.\\

Finally, we check the effect of the anharmonic coefficient $U_3$ on the thermal conductivity. We perform the calculation on the atomic chain with $N=80$ atoms and $U_4=0.008$. We vary $U_3$ and find that with the increasing of $U_3$, $\kappa$ increases slowly and then decreases rapidly after reaching a peak, as shown in Fig.3(d). The rapid decreasing of $\kappa$ with a large value of $U_3$ can be understood that the anharmonic interaction blocks the flowing of the LBs in the chain. Comparably, a small value of $U_3$ can enhance the interaction between the atoms and benefit the transport of the LBs. This result obtained in the present study is the result of the mean field, and has not been justified if the fluctuations of the thermal current are involved. \\

\section{discussion}
In the ensemble theory we proposed, the details of the interaction between the crystals and the baths are dropped off. Therefore, the properties of the thermal transport in the crystals obtained by this theory are intrinsic, and are not influenced by the details of the baths. And the calculations can be reduced from the full Hamiltonian of the whole atomic chain to the Hamiltonian of the only TAS. That means the computational loads are dependent on the calculations of two or a few atoms as an unit, which is less than the computational loads for the full atomic chain. \\

In the present study, we investigate the thermal transport in an atomic chain, which comprises identical atoms forming the periodical crystal. This study can be generalized to composite crystals. In the composite crystals, we can consider two crystal cells as one unit just like we consider two atoms as one unit in the present study. The study for the composite crystals is still under research, and not presented here.\\

The atomic chain we study is in the steady state where the thermal current is a constant and time independent. The interaction potentials between the atoms play their role to drive LBs from atom to atom, and do not absorb or release the thermal energy. Thus, our theory is a theory of the mean field by neglecting the fluctuations of the thermal current. In order to study the thermal transport much more precisely, we need go further to consider the fluctuations of the thermal current to beyond the mean filed theory, which is our future work. If the fluctuations of the thermal currents are considered, the relation between the length and $\kappa$ can be investigated precisely as well as the relation between $U_3$ and $\kappa$.\\

\section{conclusions}
Motivated by the ensemble theory of the equilibrium statistical mechanics, we propose the ensemble theory for the non-equilibrium statistics by introducing weight factors for the atoms. Our theory is used to study the thermal conductivity of the atomic chain for the demonstration. We quantize the lattice vibrations of the chain by Local Bosons instead of Phonons. We isolate two atoms as an unit from the full chain and check the time evolving of the quantum states of the two atoms. The Local Bosons flow between the two atoms, behaving like the Rabi oscillation. According to the time evolving of the quantum states of the two atoms, we define the thermal current. In this way, the properties of the thermal transport of the atomic chain can be investigated.\\

In this theory, the anharmonic potential between atoms can be fully considered without the treatment of the perturbation. We find that with the increasing of the anharmonic potential, the thermal conductivity $\kappa$ is enhanced initially and then decreases if the anharmonic potential is too large. In this study, we have not considered the fluctuations of the thermal currents, which will be our future work.\\

\begin{acknowledgments}
The author kindly acknowledges Prof. Ning-Hua Tong from Renmin University of China and Prof. Yun-An Yan from Ludong University for discussions.\\
\end{acknowledgments}

\appendix
\section{$U_2=-\omega/4$}
\label{appA}
We start from the Hamiltonian of an atomic chain with harmonic potentials, which reads
\begin{align}
\label{A1}
H=\sum_j \frac{P_j^2}{2M}+\frac{1}{2}\sum_{j,k}V_{jk}r_jr_k
\end{align}
with $P_j$ the momentum of the $j$-th atom, $r_j$ the displacement of the $j$-th atom from its equilibrium position and $V_{jk}$ the force parameter between the $j$-th and the $k$-th atoms.\\

We have defined the creation and annihilation operators in Eq.(\ref{aoperator}) to obtain
\begin{align}
\label{A2}
r_j=\sqrt{\frac{\hbar}{2\omega M}}(a_j^{\dagger}+a_j),~~~~P_j=i\sqrt{\frac{\hbar \omega M}{2}}(a_j^{\dagger}-a_j).
\end{align}
We split the second term in Eq.(\ref{A1}) into two terms by $\sum_{j,k}V_{jk}=\sum_{j}V_{jj}+\sum_{j,k\neq j}V_{jk}$, and set $V_{jj}=M\omega^2$ as we have done in the context of this paper. We substitute Eq.(\ref{A2}) into Eq.(\ref{A1}), and we have
\begin{align}
\label{A3}
H=\sum_j\hbar \omega(a_j^{\dagger}a_j+\frac{1}{2})+\frac{1}{2}\sum_{j, k\neq j}\frac{\hbar V_{jk}}{2\omega M}A_jA_k
\end{align}
with $A_j=a_j+a_j^{\dagger}$. In this study, we consider the interactions between only the nearest neighbors. Thus, the second term in the Eq.(\ref{A3}) can be simplified to be
\begin{align}
\frac{1}{2}\sum_{j, k\neq j}\frac{\hbar V_{jk}}{2\omega M}A_jA_k=\sum_{j}\frac{\hbar V_{j(j+1)}}{2\omega M}A_jA_{j+1}
\end{align}
with the factor $1/2$ dropped off for the double countering of the indexes. If the potential in the atomic chain is harmonic, we have $V_{jj}=-V_{j(j+1)}-V_{(j-1)j}$ according to the theory of the lattice dynamics. So we have $V_{j(j+1)}=-V_{jj}/2=-M\omega^2/2$. In this way, we rewrite the Hamiltonian as 
\begin{align}
H=\sum_j\hbar \omega(a_j^{\dagger}a_j+\frac{1}{2})-\frac{\hbar \omega}{4}\sum_{j}A_jA_{j+1}.
\end{align}
We have denoted the factor $-\hbar \omega /4$ by $U_2$ in this paper. By using the energy scale $\hbar \omega_0$, $U_2$ equals $-\omega/4$. In this study with the anharmonic potentials introduced for the chain, we apply $U_2=-\omega/4$ for the approximation.
\nocite{*}

\bibliography{references}% Produces the bibliography via BibTeX.

\end{document}